\documentclass{Interspeech}

\interspeechcameraready

\usepackage[T1]{fontenc}
\usepackage{subcaption}
\usepackage{multirow}

\usepackage{url}

\usepackage[labelformat=simple]{subcaption}

\title{French Listening Tests for the Assessment of Intelligibility, Quality, and Identity of Body-Conducted Speech Enhancement}

\author[affiliation={1}]{Thomas}{Joubaud}
\author[affiliation={1,2}]{Julien}{Hauret}
\author[affiliation={1}]{Véronique}{Zimpfer}
\author[affiliation={2}]{Éric}{Bavu}

\affiliation{Acoustics and Protection of the Soldier}{French-German Research Institute of Saint-Louis}{France}
\affiliation{Laboratoire de Mécanique des Structures et des Systèmes Couplés}{Conservatoire National des Arts et Metiers}{France} 
\email{\{thomas.joubaud, veronique.zimpfer\}@isl.eu, \{julien.hauret, eric.bavu\}@lecnam.net}

\graphicspath{{figures}}

\begin{document}

\maketitle

\begin{abstract}
This study evaluates the Extreme Bandwidth Extension Network (EBEN) model on body-conduction sensors through listening tests. Using the Vibravox dataset, we assess intelligibility with a French Modified Rhyme Test, speech quality with a MUSHRA (MUltiple Stimuli with Hidden Reference and Anchor) protocol and speaker identity preservation with an A/B identification task. The experiments involved male and female speakers recorded with a forehead accelerometer, rigid in-ear and throat microphones. The results confirm that EBEN enhances both speech quality and intelligibility. It slightly degrades speaker identification performance when applied to female speakers' throat microphone recordings. The findings also demonstrate a correlation between Short-Time Objective Intelligibility (STOI) and perceived quality in body-conducted speech, while speaker verification using ECAPA2-TDNN aligns well with identification performance. No tested metric reliably predicts EBEN's effect on intelligibility.

\end{abstract}

\section{Introduction}
\label{sec:intro}

Remote voice communication in noisy environments requires effective speech capture and restitution. While integrating loudspeakers into hearing protection devices generally addresses the latter, capturing speech remains challenging. Body-conduction sensors, exploiting vocal vibrations through bones and soft tissues, provide a robust alternative to traditional microphones in environments above \SI{75}{\decibel(A)} \cite{Mcbride2011effect,Bouserhal2017inear,Hauret2023configurable}. Despite their noise resilience, these sensors reduce intelligibility due to limited bandwidth. Prior research has focused on enhancing speech captured via bone-conduction transducers \cite{Tagliasacchi2020SEANet,Li2023enabling}, in-ear microphones \cite{Bouserhal2017inear, Bouserhal2015potential,Park2019speech}, and throat microphones \cite{Shahina2007mapping,Turan2013enhancement}. Recent advances leverage deep neural networks for body-conducted speech enhancement \cite{Hauret2023configurable, Tagliasacchi2020SEANet, Park2019speech,Hauret2023EBEN}. However, these approaches require extensive training data, but few publicly available datasets exist \cite{Wang2023end,esmb2021repo,Hosain2024emobone,Hauret2024vibravox}. The Vibravox dataset \cite{Hauret2024vibravox} addresses this gap by incorporating five body-conduction sensors. Using this dataset, researchers evaluated the Extreme Bandwidth Extension Network (EBEN) model \cite{Hauret2023configurable} across three tasks: speech enhancement, speech-to-phoneme transcription, and speaker verification. Objective metrics \cite{Taal2011algorithm,Manocha2022speech} confirm that EBEN improves speech quality, intelligibility, and transcription accuracy, but degrades speaker identity recognition.

Speech enhancement evaluation can rely on a wide range of metrics — 12 in \cite{Zhang2024urgent}, for instance. However, such metrics do not always align with human perception. While advancements like Audiobox \cite{tjandra2025aes} push audio evaluation boundaries, listening tests remain the gold standard. This study validates the observations from \cite{Hauret2024vibravox} through listening tests based on the Vibravox dataset. To keep test durations manageable, we focus on three body-conduction sensors: the forehead accelerometer (Knowles BU23173-000), the rigid in-ear (RIE) microphone \cite{Denk2017individualised}, and the throat microphone. An airborne headset microphone serves as a reference.
This study does not consider external noise because body-conducted sensors are inherently resilient to it. Given that their use is not necessarily limited to continuously noisy environments, our assessments with quiet recordings provide a relevant and representative setting.
In the following sections, we evaluate speech intelligibility using a French Modified Rhyme Test (MRT) \cite{Zimpfer2020development}, speech quality via a MUSHRA (MUltiple Stimuli with Hidden Reference and Anchor) test \cite{ITU2015method}, and speaker identity preservation with an A/B identification task. Additionally, we analyze results separately for male and female speakers to distinguish low- and high-pitched voices. Finally, we compare listening test outcomes with corresponding objective metrics to assess their suitability for evaluating body-conducted speech signals, both raw and enhanced.
Before conducting t-tests, we identify and remove outliers using the interquartile range (IQR) method, discarding values deviating by more than $1.5 \times IQR$ from the first or third quartile. We then verify data normality with the Shapiro-Wilk test \cite{Shapiro1965analysis}. A significance level of \SI{95}{\percent} is applied throughout our analysis.

Beyond assessing the impact of bandwidth extension, this study also contributes to a broader discussion on the relationship between objective metrics and human evaluations in speech enhancement. While automated measures are widely used for their efficiency and reproducibility, their ability to reflect perceptual quality remains an open question, particularly for body-conducted speech for which subjective studies are still scarce. By systematically comparing human judgments with metric-based assessments, our work provides insights into which metrics are best aligned with perception in this context. These findings could inform the development of more perceptually relevant evaluation frameworks for future speech enhancement systems.

\vspace*{-.3cm}

\section{Speech Intelligibility: MRT}
\subsection{Experimental Protocol}
We evaluate the intelligibility of body-conducted speech with the MRT, the standard method for communication systems according to the American National Standards Institute \cite{ANSI2009method}. The French adaptation of this test \cite{Zimpfer2020development} quantifies consonantal confusion in a closed-response set of 50 lists, each containing six Consonant-Vowel-Consonant (CVC) words. In each list, the words differ by only one consonant. To create our test material, we recruited a male and a female participant to record the full set of 300 words, using the same sensors as in \cite{Hauret2024vibravox} under quiet conditions. These participants, not included in the initial Vibravox dataset, pronounced each target word within a fixed carrier sentence: (\textit{Le mot \dots{} doit être indiqué.}\footnote{In English: \textit{The word \dots{} must be indicated.}}). The resulting MRT test data\footnote{\ifinterspeechfinal\url{https://huggingface.co/datasets/Cnam-LMSSC/french-mrt}\else Anonymous Huggingface link for review\fi} has been made publicly available on HuggingFace, along with an enhanced version\footnote{\ifinterspeechfinal\url{https://huggingface.co/datasets/Cnam-LMSSC/french-mrt_enhanced_by_EBEN}\else Anonymous Huggingface link for review\fi} processed using the appropriate EBEN models.
In this study, we focus on three body-conduction sensors: the forehead accelerometer, the rigid in-ear microphone, and the throat microphone, evaluating both their raw and enhanced signals. Including the reference microphone, this results in a total of seven test conditions. 23 native French speakers participated in the MRT experiment, split into two sessions — one for the male speaker and one for the female speaker. In each session, participants heard 50 sentences (one per MRT list) for each condition. All signals were loudness-normalized to \SI{-36}{LUFS} \cite{ITU2023algorithms}. After listening, each participant selected the target word from the six-word list. Each session lasted about 30 minutes.

\subsection{Results}
\begin{figure}[ht!]
	\centering
	\begin{subfigure}{\linewidth}
		\includegraphics[width=\linewidth]{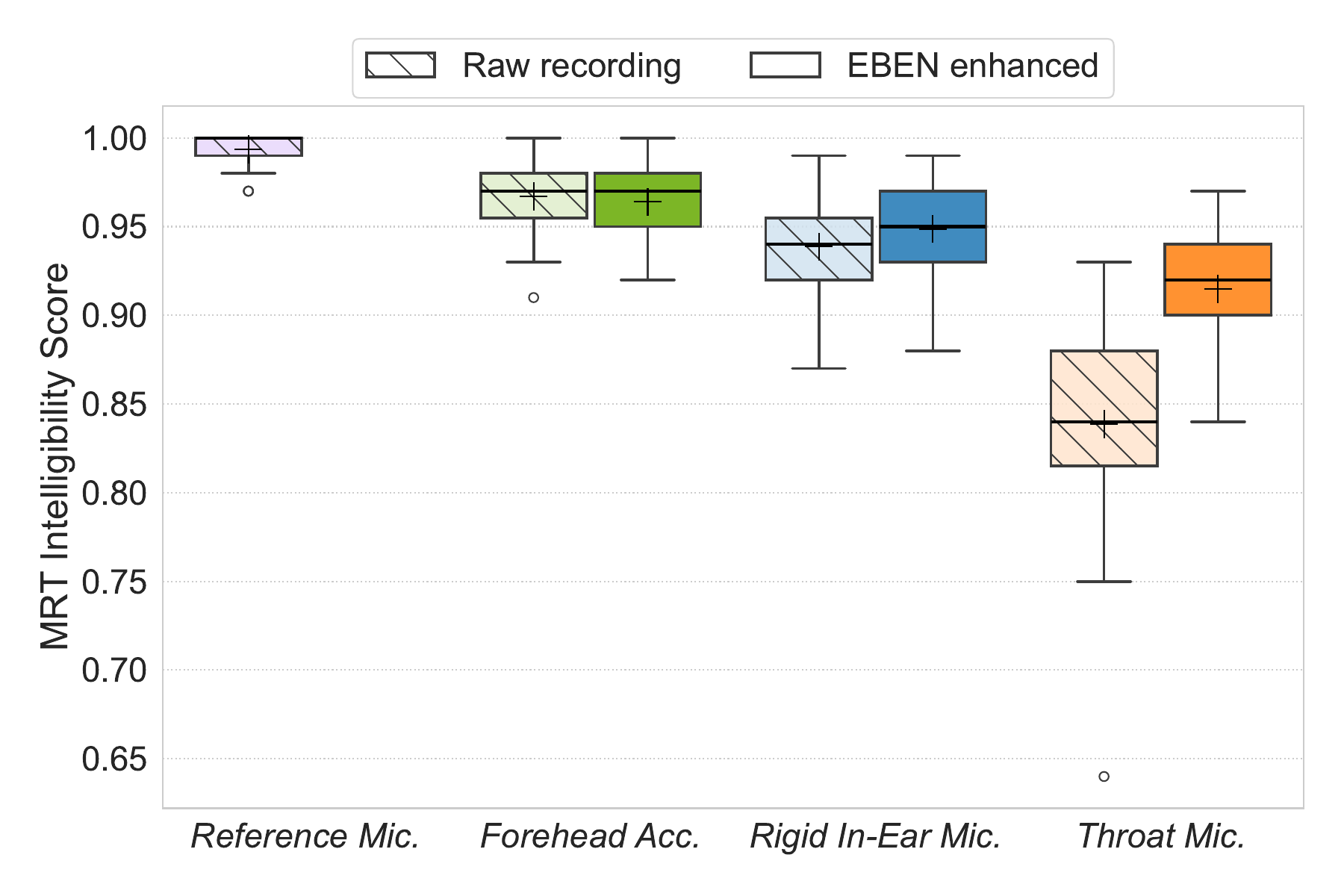}
		\caption{Distribution of MRT intelligibility score}
		\label{fig:mrt_a}
	\end{subfigure}
	\begin{subfigure}{\linewidth}
		\includegraphics[width=\linewidth]{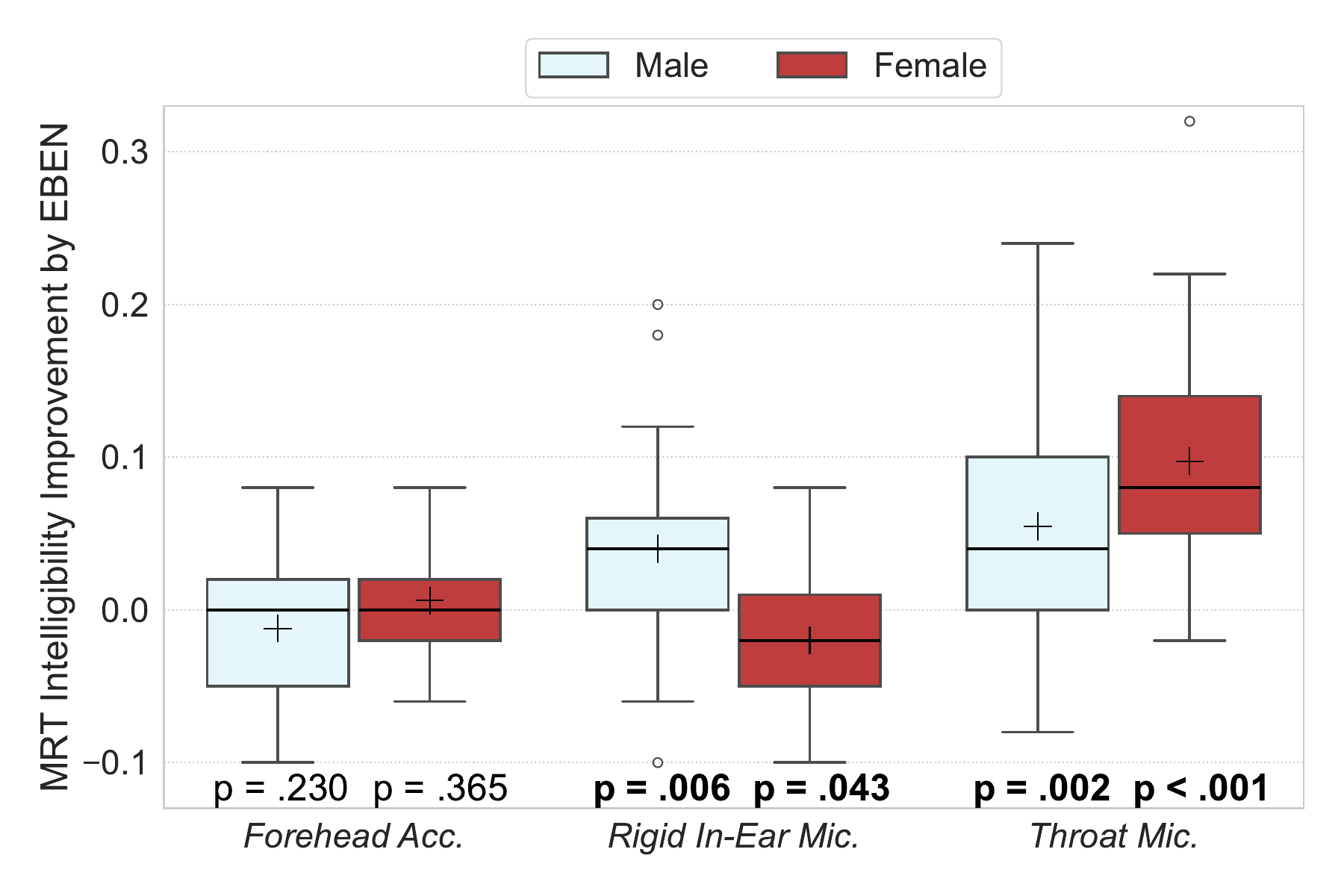}
		\caption{Distribution of EBEN-induced intelligibility improvement}
		\label{fig:mrt_b}
	\end{subfigure}
	\caption{(a) MRT intelligibility score for raw and EBEN-enhanced signals. (b) EBEN-induced intelligibility improvement per sensor and speaker gender. (Black cross: mean. p-values from one-sample t-tests for zero mean.)}
	\label{fig:mrt}
	\vspace*{-.3cm}
\end{figure}

Figure~\ref{fig:mrt_a} shows the distribution of MRT scores across all conditions. For body-conduction microphones, the average score stays above \SI{80}{\percent} for raw signals, but never matches the reference microphone's performance. EBEN has little effect on the forehead accelerometer and rigid in-ear (RIE) microphone, likely due to their high-quality raw signals. However, with the throat microphone, EBEN improves the MRT score by over \SI{5}{\percent}, demonstrating its effectiveness in this case.

For all listeners, we compute the difference between average MRT scores with EBEN-enhanced and raw signals to measure intelligibility improvement. Figure~\ref{fig:mrt_b} shows the distribution by speaker gender. As noted, EBEN has no effect on the forehead accelerometer but improves the throat microphone performance — by \SI{5}{\percent} for the male speaker and \SI{10}{\percent} for the female speaker. For the RIE microphone, there’s a \SI{4}{\percent} improvement for the male speaker but a slight \SI{2}{\percent} degradation for the female speaker. Statistical analysis supports these findings. After outlier removal and normality verification, we apply a one-sample t-test for each sensor and speaker to check if the mean intelligibility improvement differs from zero. p-values are shown in Figure~\ref{fig:mrt_b}. The female speaker's raw RIE performance (\SI{96}{\percent}) is higher than the male's (\SI{91}{\percent}), which may explain the slight degradation after enhancement.

Overall, EBEN enhances intelligibility when the body-conduction sensor introduces significant degradation, as with the throat microphone. Otherwise, its impact is minimal (forehead accelerometer) or slightly negative (RIE for the female speaker). Expanding the dataset with more speakers could confirm these trends.

\vspace*{-.2cm}

\section{Speech Quality: MUSHRA}
\subsection{Experimental Protocol}

We assess speech quality using a MUSHRA test \cite{ITU2015method} with the same body-conduction sensors and enhancement processing. From the Vibravox test set, we randomly selected 10 sentences from 5 women and 5 men. The headset microphone serves as the reference, and the temple contact microphone acts as a low-quality anchor due to its reduced bandwidth and high background noise. We evaluate raw and EBEN-enhanced signals from the forehead accelerometer, rigid in-ear microphone, and throat microphone, sourced from \footnote{\ifinterspeechfinal\url{https://huggingface.co/datasets/Cnam-LMSSC/vibravox_enhanced_by_EBEN}\else Anonymous Huggingface link for review\fi}. Since EBEN operates at 16 kHz, the raw signals are downsampled accordingly, with only the reference remaining at 48 kHz, with a downsampled hidden version. All signals are loudness-normalized to \SI{-36}{LUFS} \cite{ITU2023algorithms}. A total of 21 experienced listeners participated in the MUSHRA test. For each of the 10 sentences, they compared signals from the 8 conditions against the reference, rating speech quality on a 0–100 scale (0–20: bad, 20–40: poor, 40–60: fair, 60–80: good, 80–100: excellent). We discarded three participants who rated the hidden reference below $80$ in more than 15\% of trials. We also ensured that no listener rated the low-quality anchor as excellent in more than 15\% of cases.

\vspace*{-.2cm}
\subsection{Results}
\begin{figure}[ht!]
	\centering
	\begin{subfigure}{\linewidth}
		\includegraphics[width=\linewidth]{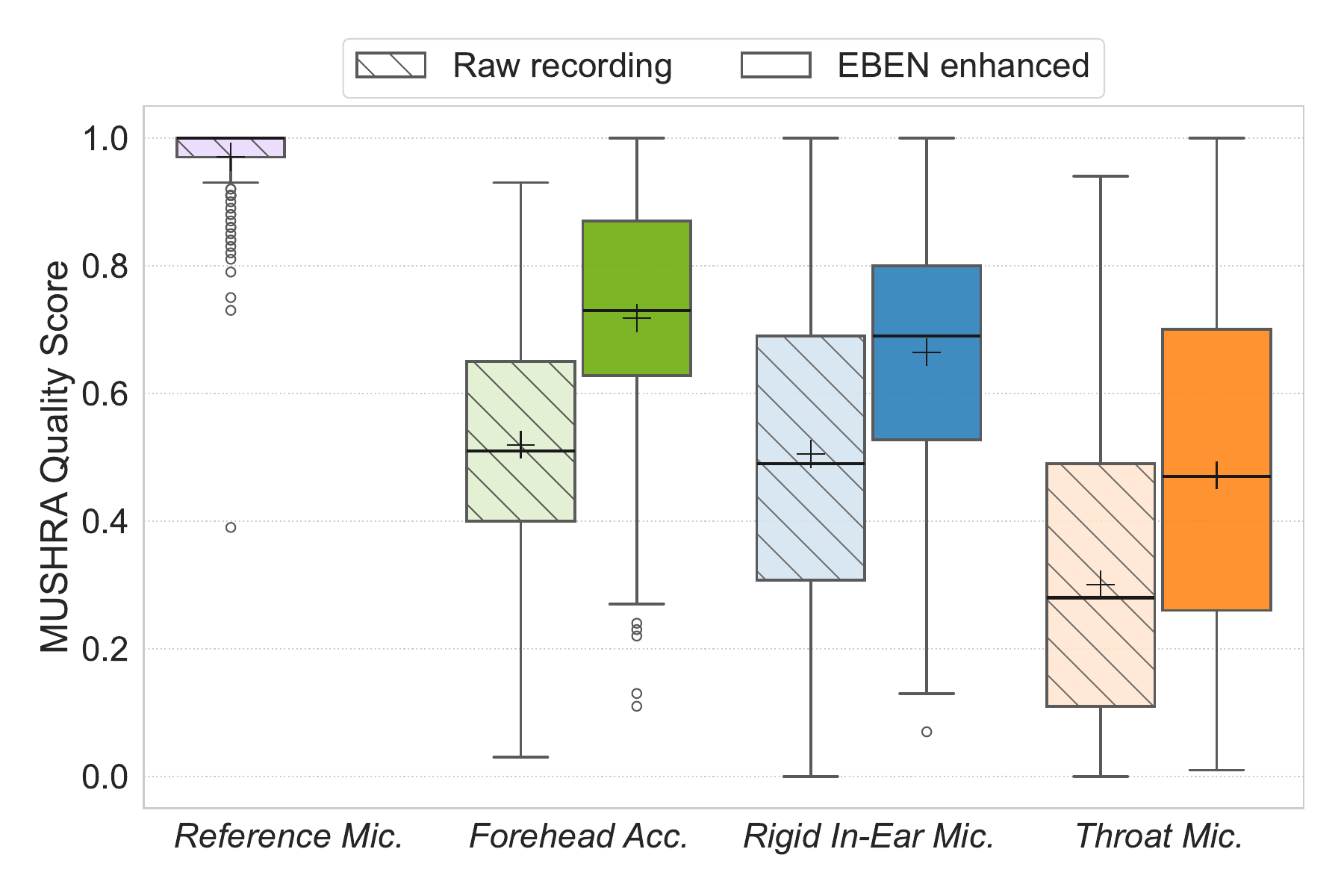}
		\caption{Distribution of MUSHRA quality rating}
		\label{fig:mushra_a}
	\end{subfigure}
	\begin{subfigure}{\linewidth}
		\includegraphics[width=\linewidth]{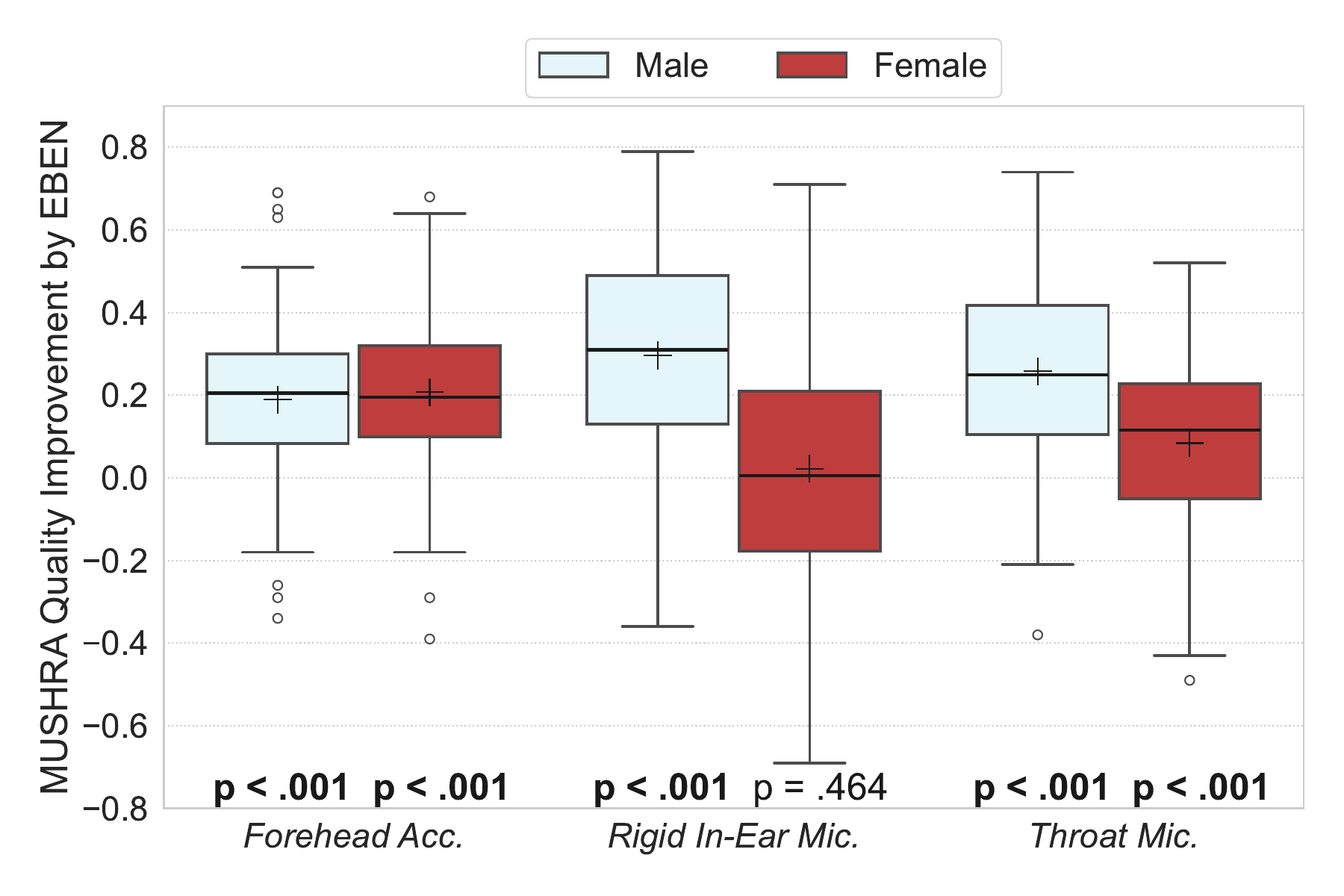}
		\caption{Distribution of EBEN-induced quality improvement}
		\label{fig:mushra_b}
	\end{subfigure}
	\caption{(a) MUSHRA quality score for raw and EBEN-enhanced signals. (b) EBEN-induced quality improvement per sensor and speaker gender. (Black cross: mean. p-values from one-sample t-tests for zero mean.)}
	\label{fig:mushra}
	\vspace*{-.1cm}
\end{figure}

Figure~\ref{fig:mushra_a} shows the quality rating distribution for all sensors, comparing raw and EBEN-enhanced signals. Listeners correctly identified the hidden reference as the best. EBEN processing improves quality for body-conduction sensors by roughly \SI{20}{\percent} on average. However, high variability in MUSHRA scores suggests differences in individual rating strategies. To address this, we compute the quality difference between EBEN-enhanced and raw signals for each listener, shown in Figure~\ref{fig:mushra_b}. A positive value indicates improvement. With the forehead accelerometer, speech quality increases by \SI{20}{\percent} on average, for both male and female speakers. For the rigid in-ear and throat microphones, enhancement benefits male speakers more significantly.

\vspace*{-.2cm}

Using one-sample t-tests (p-values in Figure~\ref{fig:mushra_b}), we find that the mean quality difference is significantly different from 0 in all conditions except for the RIE microphone with female speakers ($p=.464$). Notably, the raw signals for female speakers are rated as good on average, so the lack of improvement is not impactful. In contrast, raw (and enhanced) throat microphone signals for female speakers are rated as bad (and poor). Overall, EBEN significantly improves the quality of degraded body-conducted signals, though ratings may remain below fair (<40) for severely degraded raw signals.

\vspace*{-.1cm}
\section{Speaker Identity: A/B Identification}
\subsection{Experimental Protocol}
In \cite{Hauret2024vibravox}, the authors used a speaker verification model \cite{Thienpondt2023ecapa2} to show that EBEN-enhanced speech alters the Equal Error Rate (EER) compared to raw signals. In their study, they assessed speaker identity by comparing pairs of signals captured with the same sensor, either raw or EBEN-enhanced. To validate these findings with a listening test, we use an A/B identification approach with the Vibravox \textit{speech-clean} test set. We test the headset reference microphone, forehead accelerometer, rigid in-ear microphone, and throat microphone, in both raw and EBEN-enhanced conditions. For each test step, a sentence is randomly selected from the set, followed by another sentence either from the same speaker or a different one of the same gender. Loudness is normalized to \SI{-36}{LUFS} \cite{ITU2023algorithms}. After listening to both signals, the listener indicates whether the sentences were recorded by the same speaker. The experiment consists of 100 test steps, with 22 volunteers participating.

\vspace*{-.1cm}
\subsection{Results}
\vspace*{-.3cm}
\begin{figure}[ht!]
	\centering
	\begin{subfigure}{\linewidth}
		\includegraphics[width=\textwidth]{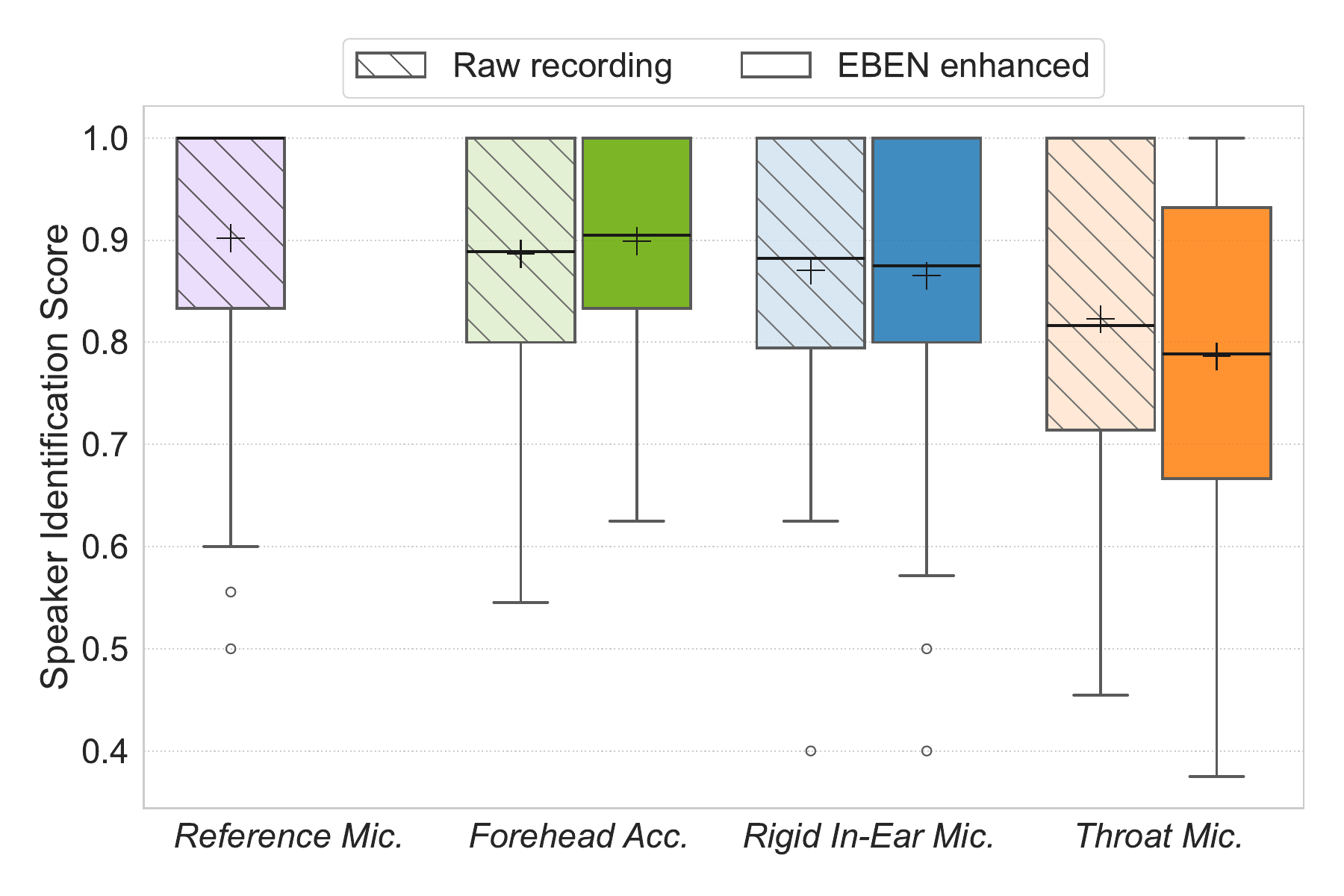}
		\caption{Distribution of A/B identification score}
		\label{fig:ab_a}
	\end{subfigure}
	\begin{subfigure}{\linewidth}
		\includegraphics[width=\textwidth]{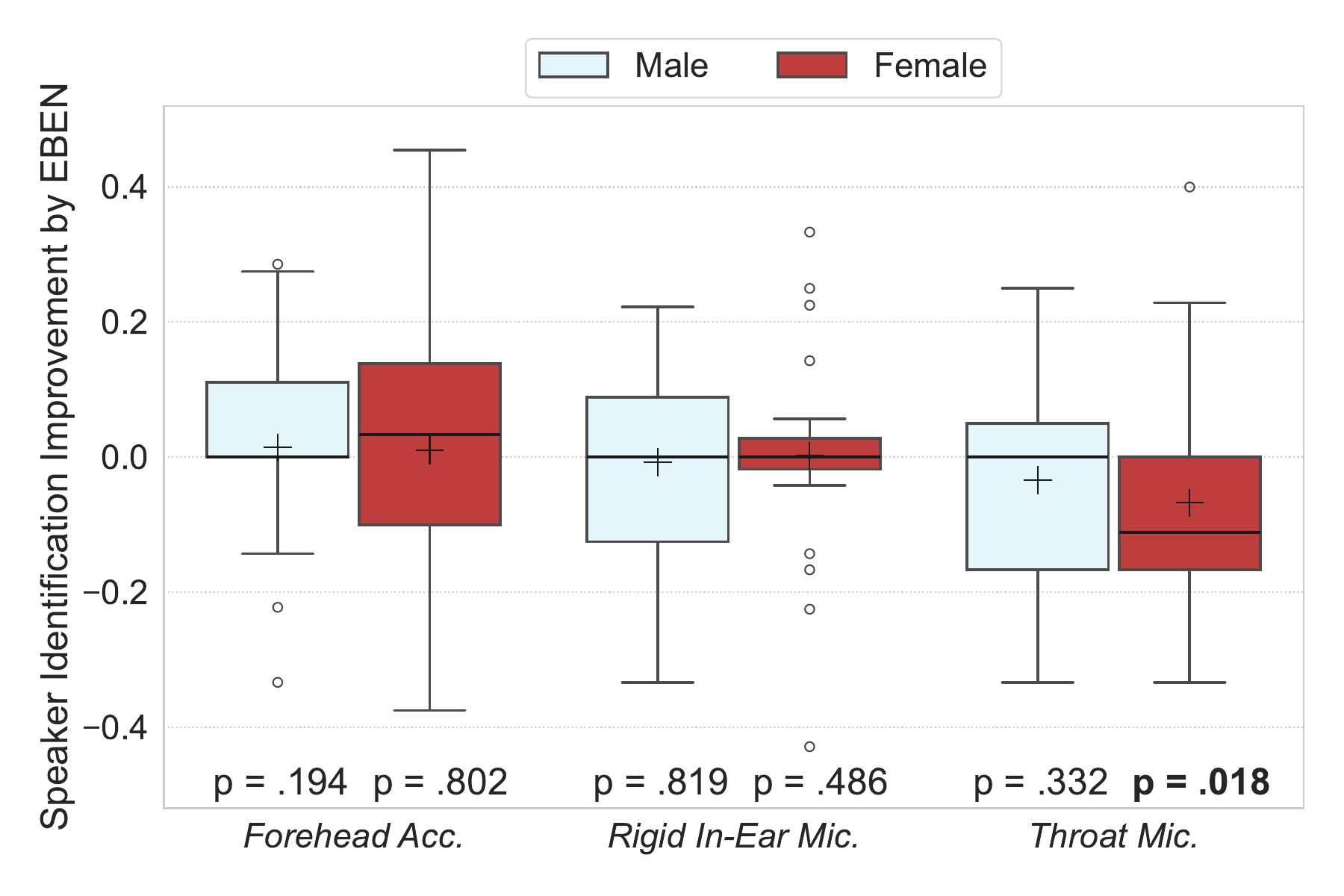}
		\caption{Distribution of EBEN-induced identification improvement}
		\label{fig:ab_b}
	\end{subfigure}
	\caption{(a) A/B speaker identification score for raw and EBEN-enhanced signals. (b) EBEN-induced identification improvement per sensor and speaker gender. (Black cross: mean. p-values from one-sample t-tests for zero mean.)}
	\label{fig:ab}
	\vspace*{-.1cm}
\end{figure}

Figure~\ref{fig:ab_a} shows the distribution of identification scores for all sensors and processing conditions. With the reference microphone, half of the listeners achieve a perfect score, and the average score is \SI{90}{\percent}. Similar performance is observed with the forehead accelerometer and RIE microphone, with no noticeable effect of EBEN processing. For the throat microphone, the mean score is \SI{82}{\percent} for raw signals, dropping \SI{4}{\percent} with EBEN.

To further assess the impact of speech enhancement on speaker identification, we compute the difference between raw and EBEN-enhanced scores and perform one-sample t-tests for zero mean. Results, shown in Figure~\ref{fig:ab_b}, indicate no significant effect for the forehead accelerometer and RIE microphone. For male speakers with the throat microphone, there is also no significant change. However, for female speakers, there is a significant reduction in identification scores, with a mean decrease of \SI{-6}{\percent} (median: \SI{-11}{\percent}).

While the EBEN model was not trained to preserve speaker identity \cite{Hauret2023configurable}, our A/B identification test shows that speech enhancement generally does not affect speaker recognition. However, results also indicate that EBEN may hinder speaker identification in cases when the original signal quality and intelligibility are poor.

\vspace*{-.2cm}
\section{Comparison with objective metrics}
In most studies related to speech enhancement, objective metrics are used to assess improvements without the need for listening tests, making it easier to compare results from different papers using the same datasets. However, the context in which these metrics have been developed may differ from the contexts in which they are applied. Our study leverages this opportunity to better understand the links and correlations between objective metrics and human evaluations for intelligibility, quality, and identity in body-conducted speech enhancement. While previous studies have attempted to align these two approaches, our work aims to provide deeper insights that can benefit the speech processing community by enhancing the understanding of how objective metrics relate to human assessments.

\vspace*{-.1cm}
\subsection{Intelligibility}
The STOI (Short-Time Objective Intelligibility) \cite{Taal2011algorithm} is one of the most widely used metrics for predicting intelligibility. We therefore compare it to the Articulation Band Correlation MRT (ABC-MRT) \cite{Voran2013using}, adapted for the MRT paradigm, and a speech-to-phone (STP) transcription model. In \cite{Hauret2024vibravox}, a Wav2Vec2.0 model \cite{Baevski2020wav2vec} was fine-tuned with the Vibravox dataset. For our analysis, we use the model trained with the reference headset microphone. Intelligibility is predicted as $1-PER$ (Phoneme Error Rate). To simulate the listening test, we computed the three metrics on the same MRT sentence recordings. The French MRT word lists target 17 consonants. We averaged the listening test performance, STOI, ABC-MRT, and STP predictions for each consonant across all raw and EBEN-enhanced sensors. The Pearson correlation coefficient $\rho$ was used to assess the metrics' suitability for predicting MRT results. As shown in Table~\ref{tab:corr}, ABC-MRT is the most correlated metric. However, $\rho$ never exceeds 0.57, highlighting the need for an intelligibility metric specifically designed for body-conducted speech. STOI fails to capture variations across consonants within a recording condition, as noted in \cite{Joubaud2024convolutional}. Lastly, STP transcription could improve by focusing only on the MRT word in the recorded carrier sentence.

\begin{table}[ht!]
	\caption{Pearson correlation coefficients between listening tests and objective metrics for intelligibility, quality, and speaker identity.}
	\label{tab:corr}
	\centering
	\begin{tabular}{llc}
		\toprule
		\textbf{Listening Test}      								& \textbf{Metric}	& \textbf{$\rho$}	\\
		\midrule
		\multirow{3}{*}{\textit{Intelligibility}}		& STOI				& .52				\\
		& ABC-MRT			& \textbf{.57}				\\
		& 1 - PER				& .45				\\
		\midrule
		\multirow{3}{*}{\textit{Quality}}				& STOI				& \textbf{.87}				\\
		& PESQ				& .81				\\
		& N-MOS				& .76				\\
		\midrule
		\textit{Identity}								& ECAPA2			& \textbf{.90}				\\
		\bottomrule
	\end{tabular}
	\vspace*{-.2cm}
\end{table}




\vspace*{-.1cm}
\subsection{Quality}
In \cite{Hauret2023configurable}, the authors found that STOI \cite{Taal2011algorithm} and N-MOS \cite{Manocha2022speech} better predicted their MUSHRA test when comparing speech enhancement models. Since STOI focuses on intelligibility, we also include wideband PESQ (Perceptual Evaluation of Speech Quality) \cite{ITU2001perceptual,ITU2007wideband} in this study. We compute the metrics on the same 10 sentences used in the MUSHRA test for all raw and EBEN-enhanced sensors. The quality section of Table~\ref{tab:corr} shows the Pearson correlation coefficients between the listening test results and predictions. All values are acceptable ($>.75$), with STOI being the best predictor ($\rho=.87$). Thus, STOI is more suitable for assessing quality than intelligibility in body-conduction sensors. In this study, PESQ is also a good indicator for quality, contrary to the findings of \cite{Hauret2023configurable}. Further training of N-MOS with body-conducted speech data could improve its predictive accuracy.

\vspace*{-.1cm}
\subsection{Identity}
Similarly to \cite{Hauret2024vibravox}, we employ a pre-trained\footnote{\url{https://huggingface.co/Jenthe/ECAPA2}} ECAPA2-TDNN model \cite{Thienpondt2023ecapa2} to extract speaker embeddings of two tested sentences. We then compute the cosine similarity between the embeddings as a prediction that the same speaker pronounced the sentences. For all raw and EBEN-enhanced sensors, we average this metric and the identification results for each of the 21 listeners of the Vibravox test set and separately if the second sentence is from the same speaker or not. The obtained Pearson correlation coefficient of .90 in Table~\ref{tab:corr} is high, confirming the metric's suitability. However, in \cite{Hauret2024vibravox}, the authors found EBEN-induced degradation of speaker identification, which we don't observe in the listening test (except for female speakers with the throat microphone). This difference may stem from human variability in listening tests, which could blur possible EBEN-induced effects.

\vspace*{-.3cm}

\section{Conclusion}
In this study, we conducted listening tests to evaluate the effectiveness of the EBEN model for body-conducted speech enhancement. When the sensor significantly degrades the signal, EBEN improves both quality and intelligibility. If the initial quality is adequate, EBEN generally maintains it, with the only exception being a slight intelligibility reduction with the RIE microphone for the tested female speaker. Moreover, despite not being explicitly trained to preserve speaker identity, EBEN does not significantly affect speaker identification, except in the case of female speakers with the throat microphone.
Lastly, we compared the listening test results with popular objective metrics. STOI and the ECAPA2 model proved to be strong predictors for speech quality and speaker identity, respectively. However, the findings suggest that other neural network-based methods, like NORESQA-MOS, would significantly benefit from training on data featuring body-conduction degradation. Additionally, prediction methods for the intelligibility of short MRT words are still underdeveloped and require further refinement to enhance their predictive accuracy for speech enhancement models. Lastly, while this study focuses on EBEN, we believe the findings generalize to a wider class of speech enhancement models. Indeed, EBEN adopts an architecture and training procedure similar to those used in widely recognized models such as Demucs \cite{Defossez2020real}, SEANet \cite{Tagliasacchi2020SEANet}, and MelGAN \cite{Kumar2019melgan}, and shares key design principles with recent neural audio codecs like SoundStream \cite{zeghidour2021soundstream}, Encodec \cite{defossez2023high} and Mimi \cite{Kyutai2024moshi}.

\bibliographystyle{IEEEtran}
\bibliography{mybib}

\end{document}